\documentclass[a4paper,10pt,aps,pre,twocolumn,amsmath,amssymb,nofootinbib]{revtex4-1}
\usepackage{graphicx,color,soul}
\usepackage[colorlinks=true,linkcolor=blue,citecolor=red]{hyperref}

\newcommand{\mean}[1]{\langle #1 \rangle}
\newcommand{\gdot}[1]{\dot\gamma #1}
\renewcommand{\vec}[1]{\mathbf #1}

\newcommand{\Gef}{G_\text{eff}}
\newcommand{\sig}{\sigma}

\newcommand{\RNum}[1]{\uppercase\expandafter{\romannumeral #1\relax}}

\DeclareMathOperator{\erf}{erf}

\begin{document}

\title{Classical Nucleation Theory for the Crystallization Kinetics in Sheared Liquids}

\author{David Richard}
\altaffiliation[Present affiliation: ]{University of Amsterdam}
\affiliation{Institut f\"ur Physik, Johannes Gutenberg-Universit\"at Mainz, Staudingerweg 7-9, 55128 Mainz, Germany}
\author{Thomas Speck}
\affiliation{Institut f\"ur Physik, Johannes Gutenberg-Universit\"at Mainz, Staudingerweg 7-9, 55128 Mainz, Germany}

\begin{abstract}
  While statistical mechanics provides a comprehensive framework for the understanding of equilibrium phase behavior, predicting the kinetics of phase transformations remains a challenge. Classical nucleation theory (CNT) provides a thermodynamic framework to relate the nucleation rate to thermodynamic quantities such as pressure difference and interfacial tension through the nucleation work necessary to spawn critical nuclei. However, it remains unclear whether such an approach can be extended to the crystallization of driven melts that are subjected to mechanical stresses and flows. Here, we demonstrate numerically for hard spheres that the impact of simple shear on the crystallization rate can be rationalized within the CNT framework by an additional elastic work proportional to the droplet volume. We extract the local stress and strain inside solid droplets, which yield size-dependent values for the shear modulus that are about half of the bulk value. Finally, we show that for a complete description one also has to take into account the change of interfacial work between the strained droplet and the sheared liquid. From scaling reasons, we expect this extra contribution to dominate the work formation of small nuclei but become negligible compared to the elastic work for droplets composed of a few hundreds particles.
\end{abstract}

\maketitle


\section{Introduction}

Classical nucleation theory (CNT) is an establish thermodynamic framework that helps understanding phase formation and interpreting experiments and numerical simulations. Notable examples are the modeling of ice nucleation rates (impacting the understanding of our climate~\cite{zobrist2007heterogeneous,hoose2010classical,laksmono2015anomalous}) and the estimation of interfacial tension~\cite{knott2012homogeneous,sanz2013homogeneous,espinosa2016seeding,espinosa2016interfacial}. Originally developed to study systems that are prepared in thermal equilibrium, there have been several attempts to extend CNT to systems driven into a non-equilibrium steady state~\cite{desre1990suppression,kashchiev1972influence,reguera2003homogeneous1,reguera2003homogeneous2,mokshin2013extension,mura2016effects}. Of particular interest is crystallization in the presence of mechanical stresses and flows~\cite{reguera2003homogeneous2,mokshin2013extension,mura2016effects}. How such driving forces can control not only the nucleation kinetics but also the structure of the newborn solid phase remains poorly understood.

Already for relatively simple models, such as liquids and colloidal suspensions in which particles interact through soft or hard-core repulsions, the effect of flow on the nucleation kinetics is far from trivial and heavily depends on the shear strength as well as the degree of supersaturation (or cooling)~\cite{ackerson1981shear,haw1998direct,blaak2004crystal,holmqvist2005crystallization,cerda2008shear,wu2009melting,lander2013crystallization,shao2015shear,mao2015stress,ruiz2018crystal}. Aside from shear-induced order (particle layering) at high strain rates~\cite{ackerson1988shear,xue1990shear,panine2002structure,nikoubashman2011cluster,besseling2012oscillatory} one finds that supercooled liquids crystallize basically via the same activated nucelation process as in the quiescent regime: The system remains in the disordered melt until a rare fluctuation leads to a sufficiently large \emph{critical nucleus} that grows spontaneously. This suggests to write the crystallization rate as $k=\kappa e^{-W}$ with work $W$ (in units of the thermal energy $k_BT$) to escape from the metastable state and kinetic prefactor $\kappa$. The theoretical challenge is to find expressions for $W$ (and $\kappa$) as a function of the relevant system parameters.

In a pioneering numerical work~\cite{blaak2004crystal,blaak2004homogeneous} studying the crystallization of colloidal particles it has been shown that shear flow can suppress crystallization. This has been rationalized through a significant increase of $W$ as a function of the strain rate. However, other experimental and numerical studies have reported the opposite behavior with shear induces order \cite{holmqvist2005crystallization,mokshin2010crystal,mokshin2013extension,wu2009melting,shao2015shear} and even an optimal strain rate for which the crystallization kinetics is fastest~\cite{cerda2008shear,mokshin2013extension,lander2013crystallization}. More recently, we have probed the crystallization of hard spheres as a function of the strain rate and the packing fraction~\cite{richard2015role}, the latter serving as a control parameter for the degree of supersaturation. We found a crossover from shear-induced suppression to shear-induced enhancement: At low packing fractions, the rate is dominated by the nucleation barrier tending to increase with the strength of the shear flow. As for soft spheres, small clusters are more likely to dissolve and one finds larger critical nucleus sizes. In contrast, for dense suspensions the (quiescent) nucleation barrier effectively vanishes and the rate is controlled by the kinetic prefactor $\kappa$. The latter strongly follows the particle dynamics and drops close to the glass transition due to caging effects. In this regime external flow enhances the particle diffusion and allows for a better exploration of configuration space, accelerating the formation and growth of solid nuclei. Continuing to increase the strain rate, clusters start to break-up and the rate again is controlled by the nucleation work of critical nuclei.

In this paper, we numerically study the extension of CNT to sheared liquids. As a well-studied testbed, we choose monodisperse hard spheres. Recently, Mura and Zaccone have put forward the idea that the quiescent nucleation work is to be augmented by a reversible elastic work to stress the critical nucleus~\cite{mura2016effects}. Such an extension has already been proposed to model the steady coexistence of a strained solid with its sheared melt~\cite{butler2002factors,butler2003simulation}, but was found inconclusive with respect to the existence of a chemical potential out of equilibrium. Here, we numerically test this scenario and demonstrate that the change of nucleation kinetics of sheared hard spheres can indeed be captured through an additional elastic work. Crucially, for quantitative predictions one needs to take into account the diminished density and the elevated shear stress inside finite solid droplets.

The manuscript is organized as follows: We first discuss the theoretical basis for extending CNT in Sec.~\ref{sec:theory} before providing details of the model system and the simulations in Sec.~\ref{sec:meth}. In Sec.~\ref{sec:results}, we first show that employing bulk values for the shear modulus and equating the solid stress with the hydrodynamic liquid stress does not yield reasonable predictions. We then present a method to access the local shear stress and shear strain of finite solid droplets, which gives access to the shear modulus and shows that the stress inside the droplet is elevated. Finally, we reconstruct the elastic work as a function of the droplet size and highlight the presence of an extra surface work for small droplets.


\section{Theory}
\label{sec:theory}

\subsection{Classical nucleation theory under shear}

\begin{figure}[b!]
  \includegraphics[scale=1.0]{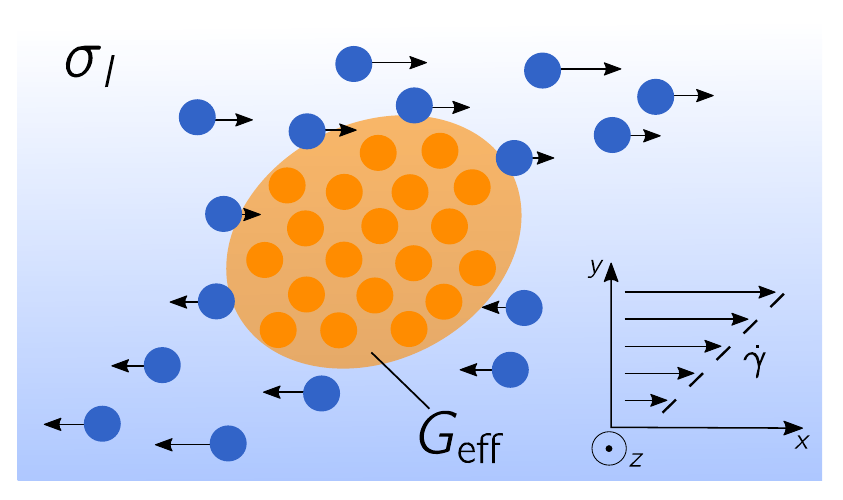}
  \caption{\textbf{Solid-liquid coexistence under shear.} Sketch of the formation of a solid droplet within the sheared melt. We distinguish in orange ``solid-like'' particles that have a high bond-order symmetry from liquid particles. For the thermodynamic modeling, we employ Gibbs concept of a sharp dividing surface separating the homogeneous droplet (orange area) from the surrounding liquid. The inset shows the simple-shear flow geometry with strain rate $\gdot$ used in our simulations.}
  \label{fig:sketch}
\end{figure}

We aim to predict the isothermal nucleation rate $k$ per volume with which a critical solid droplet spontaneously forms within the sheared melt (Fig.~\ref{fig:sketch}). The two main control parameters are the liquid density $\rho_l$ and the imposed strain rate $\gdot$. The latter determines the shear stress $\sigma_l=\eta\gdot$ in the liquid acting upon the droplet, with $\eta(\gdot,\rho_l)$ denoting the shear viscosity of the liquid at a particular density. Note that we explicitly take into account a dependence of $\eta$ on $\gdot$ since dense liquids might undergo shear thinning. The solid droplet is characterized by its volume $V_s$, its density $\rho_s(V_s)$, and its shear modulus $G(V_s)$, both of which depend on the droplet size. Throughout, we follow Gibbs idea of a dividing surface separating the solid droplet from the liquid, both of which are modeled as homogeneous systems.

Without shear flow ($\dot\gamma=0$), the nucleation rate is dominated by the reversible nucleation work to reach the transition state, \emph{i.e.}, a critical solid droplet of volume $V_s$. This nucleation work is given by the free energy difference
\begin{equation}
  \label{eq:work}
  \Delta F_0(V_s) = -\Delta P V_s + \Phi(V_s).
\end{equation}
The first term is proportional to the volume and captures the free energy gained through creating space for the droplet. The second term $\Phi(V_s)=\Gamma A$ is the excess interfacial free energy given as the product of the droplet surface $A\sim V_s^{2/3}$ and the interfacial tension $\Gamma$. In principle, $\Gamma$ again depends on the droplet size~\cite{richard2018crystallization2}. The thermodynamic driving force for nucleation is the difference $\Delta P=P_s-P_l$ of pressure $P_s$ between the inside of the solid droplet and the ambient liquid pressure $P_l$ at the same liquid chemical potential, $\mu_s=\mu_l$. It is assumed that the surrounding liquid is stress-free and therefore there is no elastic work on the nucleus entering the nucleation work.

Turning on the shear flow with $\gdot\neq 0$, the system is steadily driven away from equilibrium into a non-equilibrium steady state and thus constantly dissipates heat. Strictly speaking, there is no thermodynamic potential anymore that determines the behavior of the system. To proceed, we make the three following assumptions: First, in our modeling we neglect the dissipation due to shearing the liquid. We treat the droplet as an inclusion in a stressed medium (the surrounding liquid) but ignore the ``housekeeping'' work that needs to be spend to keep the medium at a given shear stress $\sig_l$~\cite{speck09,gerloff18}. We do, however, consider the \emph{excess work} $W$ required to form the critical solid droplet. Second, we assume that the nucleation rate is still determined by this excess work, $k=\kappa e^{-W}$, \emph{i.e.}, the droplet still emerges due to a spontaneous thermal fluctuation. Put differently, the liquid acts as a heat reservoir, the fluctuations of which are still characterized by a (possibly effective) temperature. In contrast to the quiescent liquid, however, there is an additional elastic work
\begin{equation}
  \label{eq:elwork}
  W_e(V_s) = \frac{\sigma_s^2}{2G(V_s)}V_s
\end{equation}
to create the droplet with shear stress $\sig_s$~\cite{landau1959course,butler2003simulation,mura2016effects}. Note that this work is not compensated by a reduction in free energy of the liquid since the external work to maintain $\sig_l$ is immediately dissipated. Hence, the total nucleation work now reads $W=\Delta F_0+W_e$. Third, we assume that the droplet undergoes a pure shear transformation with strain $\gamma_s=\sigma_s/G$.

We now restrict ourselves to spherical droplets with radius $R$, volume $V_s=\frac{4\pi}{3}R^3$, and area $A=4\pi R^2$. The mechanical equilibrium condition ($\partial W/\partial R=0$) yields
\begin{equation}
  \Delta P - \frac{\sigma_s^2}{2\Gef} = \frac{2\Gamma_\ast}{R_\ast} 
\end{equation}
at the surface of tension $R_\ast$ defined through $\partial\Gamma(R)/\partial R|_{R_\ast}=0$, where $\Gamma_\ast=\Gamma(R_\ast)$ and the shear modulus $G_\ast=G(R_\ast)$ are evaluated at the critical droplet radius $R_\ast$. Here we have introduced the effective shear modulus
\begin{equation}
  \Gef = G_\ast\left[ 1-\frac{R_\ast}{3G_\ast}
    \left.\frac{\partial G}{\partial R}\right|_{R_\ast}\right]^{-1}.
\end{equation}
Eliminating volume and area of the critical droplet, the nucleation work then takes the customary CNT form
\begin{equation}
  \label{eq:newwork}
  W(\sig_s;R_\ast) = \frac{16\pi\Gamma_\ast^3}{3(\Delta P-\frac{\sigma_s^2}{2\Gef})^2}
\end{equation}
known from quiescent nucleation but with $\Delta P$ replaced by the effective driving force $\Delta P-\frac{\sigma_s^2}{2\Gef}$, which is reduced due to the additional elastic work required to deform the solid nucleus.

For small shear stress $\sigma_s$ in the linear response regime, we expand the nucleation work in the form $W(\sigma_s)\approx\Delta F_0(1+a_W\sigma_s^2)$, where the CNT prediction for the response coefficient $a_W$ depends on the liquid and solid properties through $\Delta P$ and $\Gef$ as
\begin{equation}
  \label{eq:coef1}
  a_W^\text{CNT} = \frac{1}{\Gef\Delta P}.
\end{equation}
The same expansion can be performed for the critical nucleus size $N_c=\tfrac{4\pi}{3}\rho_sR_\ast^3\approx N_0(1+a_N\sigma_s^2)$ and the Zeldovich factor $Z_c=\sqrt{\frac{|F''(q_c)|}{2\pi}}\approx Z_0(1+a_Z\sigma_s^2)$, where $N_0$ and $Z_0$ are the critical nucleus size and Zeldovich factor in the quiescent limit $\gdot=0$, respectively. Moreover, for the rate we find $\ln k\approx\ln k_0+a_k\sigma_s^2$ with $k_0$ the nucleation rate at vanishing stress. From the CNT expression for the work Eq.~\eqref{eq:newwork}, we thus find the predictions
\begin{gather}
  \label{eq:coef2}
  a_N^\text{CNT} = \frac{3}{2}a_W^\text{CNT}, \\ a_Z^\text{CNT} = -a_W^\text{CNT}, \\ a_k^\text{CNT} = -\Delta F_0a_W^\text{CNT}.
\end{gather}
For notational convenience, we drop the subscript ``eff'' for the shear modulus in the following.


\section{Methods}
\label{sec:meth}

\subsection{Model}

We perform non-equilibrium Brownian dynamics (BD) simulations of a mono-component hard-sphere fluid. Particles interact through the pairwise Weeks-Chandler-Anderson potential $u(r)=4\epsilon[(\alpha/r)^{12}-(\alpha/r)^6+1/4]$ for $r<2^{1/6}\alpha$. Simulations are done in the canonical ensemble (NVT) with a fixed number of particles $N$, volume $V$, and temperature $T$. The system is composed of $N=5000$ particles if not mentioned otherwise.

As indicated in Fig.~\ref{fig:sketch}, the direction of the shear flow is set along $\vec{e_x}$, the flow gradient and vorticity along $\vec{e_y}$ and $\vec{e_z}$, respectively. The coupled equations of motion read
\begin{equation}
\dot{\vec r}_i = -\frac{D_0}{k_BT}\nabla_i U + \gdot y_i\vec{e_x}+\sqrt{2D_0}\vec \xi_i
\end{equation}
with $D_0$ the bare translational diffusion coefficient, $\vec \xi_i$ Gaussian white noise, and $-\nabla_i U$ is the conservative force acting on particle $i$. Additionally, we employ Lees and Edwards periodic boundary conditions \cite{allen2017computer}. The potential strength $\epsilon$ is set to $40k_BT$, with $k_B$ the Boltzmann constant. Throughout all our simulations, we scale lengths by $\alpha$, times by $\alpha^2/D_0$, and energies by $k_BT$. The equations of motion are integrated with time step $10^{-5}$. All results are then expressed in hard-sphere units where lengths are measured in units of the effective diameter $d=1.097\alpha$ and the Brownian time is defined as $\tau_B=d^2/D_0$. The packing fraction $\phi$ of the system is given by $\phi=\pi Nd^3/(6V)$. Further details concerning this mapping can be found in Refs.~\cite{filion2011simulation,richard2018crystallization}.

\subsection{Bond-orientational order}
\label{sec:bond}

We monitor the degree of crystallinity in our system through the local bond-orientational order parameter~\cite{Steinhardt1983,Dell08}
\begin{equation}
  \label{eq:order}
  q_{l,m}(i) = \frac{1}{N_n(i)}\sum_{j=1}^{N_n(i)} Y_{l,m}(\theta_{i,j},\varphi_{i,j}),
\end{equation}
which is evaluated for particle $i$, where $Y_{l,m}(\theta,\varphi)$ are spherical harmonics and $N_n$ is the number of neighbors within distance $r_{ij} < 1.5\alpha$. We construct a bond network through the scalar product
\begin{equation}
  \label{eq:scalar}
  d(i,j) = \frac{\sum_{m=-l}^{l} q_{l,m}(i) q_{l,m}^\ast(j)}{(\sum_{m=-l}^{l} |q_{l,m}(i)|^2)^{1/2} (\sum_{m=-l}^{l} |q_{l,m}(j)|^2)^{1/2}}
\end{equation}
using $l=6$ with $d(i,j) > 0.7$ defining a bond. Finally, a particle is defined as ``solid-like'' if the number of bonds $\xi\ge 9$, and clusters are constructed from mutually bonded solid-like particles.

\subsection{Shear viscosity and shear modulus}

\begin{figure}[t]
  \includegraphics[scale=1.0]{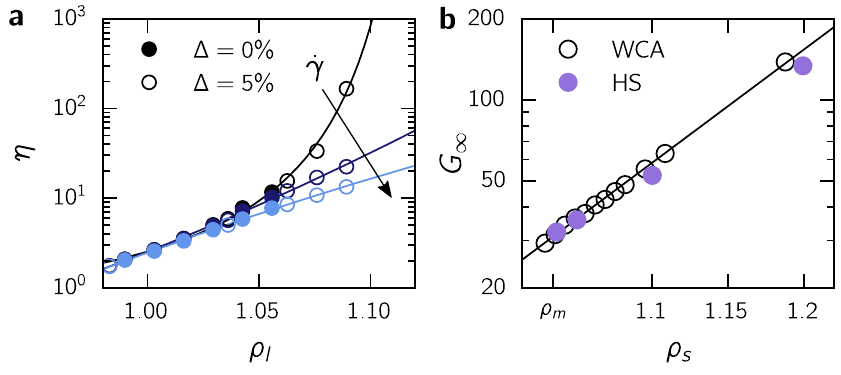}
  \caption{\textbf{Shear viscosity and shear modulus.} (a)~Shear viscosity $\eta$ as a function of the liquid density $\rho_l$ for $\dot\gamma=0.0,0.036$, and $0.084$. The solid line for $\dot\gamma=0$ is a Vogel-Fulcher-Tammann fit \cite{debenedetti2001supercooled}. Solid lines for $\dot\gamma>0$ are guides to the eye. (b) Bulk shear modulus $G_\infty$ as a function of the solid density $\rho_s$ for the WCA solid (empty black circles) and for true hard spheres (filled purple circles) taken from Ref.~\citenum{pronk2003large}. The solid line is a linear regression of $\ln G_\infty$.}
  \label{fig:visco}
\end{figure}

In a homogeneous liquid, we monitor the fluctuating shear stress $\hat\sigma$ at constant strain rate $\dot\gamma$ and deduce the shear viscosity $\eta=\langle\hat\sigma\rangle/\dot\gamma$, where the average $\langle\cdots\rangle$ here involves only configurations that have less than 5\% ``solid-like'' particles. Since dense suspensions undergo shear thinning at finite strain rates, we first evaluate $\eta$ as a function of $\dot\gamma$ and fit the flow curve using a Carreau model~\cite{bruus2008theoretical}, giving us access to the zero shear rate viscosity $\eta_0$. Beyond $\rho_l\simeq1.06$ (and for $N=5000$), monodisperse hard spheres crystallize within $<10\tau_B$, preventing the correct estimation of $\eta$. To circumvent this limitation, we have performed additional simulations with a $\Delta=5\%$ Gaussian polydispersity. Results are shown in Fig.~\ref{fig:visco}(a) for $\dot\gamma=0.0,0.036$, and $0.084$. At vanishing stress $\dot\gamma\to0$, the viscosity diverges approaching $\rho_l\simeq1.1$ ($\phi\simeq0.576$) marking the onset of dynamical arrest, which is well modeled by a Vogel-Fulcher-Tammann (VFT) form~\cite{debenedetti2001supercooled}. For $\dot\gamma>0$, we observe a drop of the viscosity indicating shear thinning.

To investigate the role of an elastic work, we require access to the stress $\sig_s$ and the shear modulus $G$ inside critical droplets as a function of droplet size $R_\ast$. As reference, we first determine the \emph{bulk shear modulus} $G_\infty=G(R_\ast\to\infty)$ as a function of the solid density $\rho_s$. To this end, we compute the bulk shear modulus of a face-centered-cubic (fcc) crystal closely following Ref.~\citenum{palmer2017thermodynamic}. In these simulations, particle positions are subjected to the affine transformation
\begin{equation}
  \vec r_i^\gamma = \vec r_i+\gamma y_i \vec{e_x}
\end{equation}
with strain $\gamma$. The shear modulus is computed from the initial slope of the average stress $\langle\hat\sigma\rangle_{\gamma}$ plotted against the imposed strain $\gamma$. In Fig.~\ref{fig:visco}(b), we compare our shear modulus with values of true hard spheres. We find a very good agreement between the two systems, supporting the validity of the mapping procedure.


\section{Results}
\label{sec:results}

\subsection{Extracting nucleation rate and work}

\begin{figure}[t!]
  \includegraphics[scale=1.0]{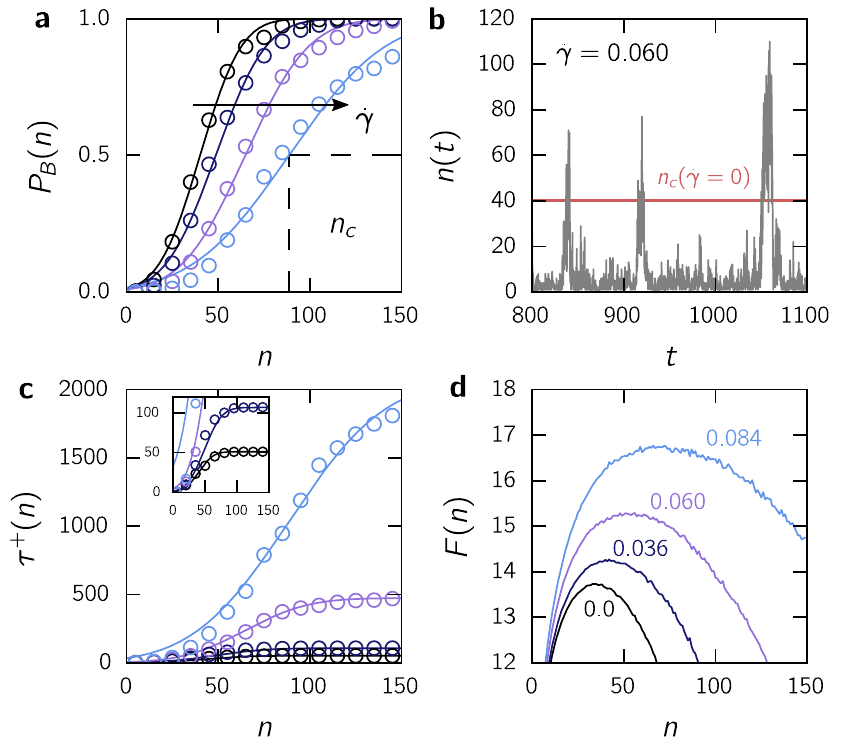}
  \caption{\textbf{Extracting kinetics and free energy barriers.} (a)~Splitting probability $P_B$ as a function of the nucleus size $n$ for various imposed shear rates $\dot\gamma$. Lines are fits to Eq.~\eqref{eq:pb}. (b)~Time evolution of the largest nucleus $n(t)$ showing multiple unsuccesful nucleation events for $\dot\gamma \simeq0.06$. The red horizontal line indicates the quiescent critical nucleus size $n_c(\dot\gamma=0)\simeq40$. (c)~Mean first passage time as a function of $n$. Solid lines are the model functions from (a) but scaled by the nucleation time $\tau_x=1/j$. The inset shows a zoom for data at small strain rates. (d)~Free energy reconstruction using $P_B$ (cf. main text). Simulations in this figure are performed at $\phi\simeq0.542$}
  \label{fig:methods}
\end{figure}

We employ the same framework as presented in Ref.~\citenum{richard2018crystallization}. We prepare initial configurations at packing fraction $\phi$ without any solid particles using the algorithm developed by Clarke and Wiley~\cite{Wiley87}, where the non-overlapping distance between particles is chosen to be equal to the effective diameter $d_\text{eff}$. We then harvest $300$ independent trajectories. As order parameter we employ the size $n$ of the largest cluster, and trajectories are terminated when the system crosses the barrier and reaches an absorbing boundary at $n_b$. Specifically, we choose $n_b=400$, which is four times larger than the largest critical size $n_c $ found in this work. This construction enables to sample a non-equilibrium steady-state with a net current $j$ of droplets flowing from the metastable melt towards $n_b$. As shown in Refs.~\cite{manuel2015reconstructing,richard2018crystallization} and below, it allows to consistently extract nucleation barriers $F(n)$ through linking the non-equilibrium distribution $P^+(n)$ and the splitting probability $P_B(n)$ that a configuration at $n$ will commit to the solid phase and reach $n_b$. For large barriers, a quadratic expansion of the free energy yields the expression
\begin{equation}
  \label{eq:pb}
  P_B(n)=\frac{1}{2}(1+\erf[\sqrt{\pi}z_c(n-n_c)]),
\end{equation}
which is used to extract critical nuclei sizes $n_c$ and Zeldovich factors $z_c$. Note that we explicitly distinguish the variables $N_c$ and $Z_c$ appearing in the CNT expressions from $n_c$ and $z_c$. The latter are computed using the bond-order parameter described in Sec.~\ref{sec:bond} and thus depend on the set of parameters employed to construct the bond network.

Fig.~\ref{fig:methods}(a) shows how $P_B$ changes with the imposed strain rate $\dot\gamma$ at $\phi\simeq0.542$, which is close to the melting point located at $\phi_m\simeq0.543$. Progressively increasing the strain rate $\dot\gamma$, we first observe a shift of $P_B$ towards larger $n$, indicating that small clusters are likely to dissolve under shear. Second, we find a systematic broadening of the splitting probability, implying a flattening of the barrier at $n\simeq n_c$, \emph{i.e.}, a smaller Zeldovich factor. Both behaviors are in qualitative agreement with Eq.~\eqref{eq:coef2}. In Fig.~\ref{fig:methods}(b), we show at $\dot\gamma \simeq0.06$ consecutive unsuccessful nucleation events, whereby nuclei larger than the quiescent critical nucleus size $n_c(\dot\gamma=0)$ fully dissolved. Furthermore, we have computed the mean first passage time (MFPT) $\tau^+(n)$ to reach a given size $n$ starting from the metastable liquid. The MFPT is inversely proportional to $P_B(n)$ scaled by the nucleation time $\tau_x=1/j$. In Fig.~\ref{fig:methods}(b), we confirm this connection by comparing $\tau^+(n)$ with $P_B(n)$ scaled by $\tau_x$, which is the value at which $\tau^+(n)$ plateaus. 

Having collected a set of trajectories, we can now compute the stationary probability distribution $P^+(n)$ to observe a configuration with a droplet of size $n$. As shown in Refs.~\cite{manuel2015reconstructing,richard2018crystallization}, we can reconstruct the actual distribution $P(n)=P^+(n)/[1-P_B(n)]$. In our simulations, we also compute the average number of clusters of size $n$, which allows to correct $P(n)$ for small clusters~\cite{richard2018crystallization}. We interpret $F(n)\sim\ln P(n)$ as an effective free energy governing the nucleation kinetics, from which we extract the nucleation work $W=\Delta F$ as the height of the barrier. Many more details of this procedure can be found in Refs.~\cite{lundrigan2009test,leitold2016nucleation,richard2018crystallization}. In Fig.~\ref{fig:methods}(d), we plot the profile $F(n)$ for several strain rates $\gdot$ at the same packing fraction. We observe that the top of the barrier moves to larger values $W$ and larger critical sizes $n_c$, and that $F(n\simeq n_c)$ flattens in agreement with a decrease of the Zeldovich factor seen from the broadening of $P_B$ and $\tau^+$. For the largest strain rate, we find a barrier increase of about $3k_BT$, which is consistent with the two orders of magnitude increase in the nucleation time seen in Fig.~\ref{fig:methods}(c).

\subsection{Response coefficients based on liquid shear stress and CNT predictions}
\label{sec:coef}

We can now extract the nucleation work $W$ and nucleation rates $k=1/(\tau_xV)$ as well as critical sizes $n_c$ and Zeldovich factors $z_c$. For moderate shear rates, plotting these quantities as a function of the square of the solid shear stress $\sigma_s^2$ would disclose the derived response coefficients in Eq.~\eqref{eq:coef1} and Eq.~\eqref{eq:coef2}. Unfortunately, defining and measuring a mechanical shear stress within finite droplets is non-trivial. As a first step to assess the impact of shear, we will plot results as a function of the ambient liquid stress $\sig_l=\eta\gdot$. The relation $\sig_s=\sig_l$ has been proposed in Ref.~\citenum{mura2016effects} but we will show below that it does not hold for finite droplets.

\begin{figure}[t]
  \includegraphics[scale=1.0]{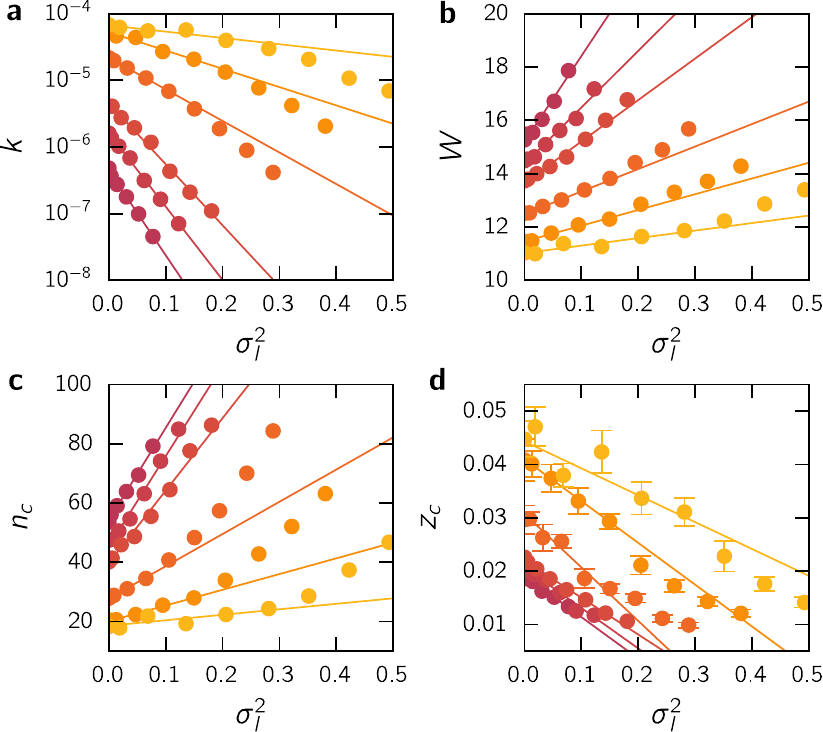}
  \caption{\textbf{Extracting response coefficients.} (a)~Nucleation rate $k$, (b)~nucleation barrier $W$, (c)~critical nucleus size $n_c$, (d)~and Zeldovich factor $z_c$ as functions of the squared shear stress $\sigma_l^2$ in the ambient liquid. Response coefficients $\{a_k,a_W,a_n,a_z\}$ are extracted from the lines, which are linear regressions only taking into account data with $\sigma_l^2<0.15$. Packing fractions are ranging from $0.539$ (dark red) to $0.553$ (light orange).}
  \label{fig:rate}
\end{figure}

As already located in our previous work~\cite{richard2015role}, we expect a change from crystallization suppression to enhancement around $\phi\simeq0.56$. Here, we focus on the regime where shear suppresses crystallization and have applied our methodology for various packing fractions ranging from $\phi\simeq0.539$ to $\phi\simeq 0.553$. In Fig.~\ref{fig:rate}, we plot $k$, $W$, $n_c$, and $z_c$ against $\sigma_l^2$. As we increase $\sigma_l$, both the rate and Zeldovich factor decrease, whereas the nucleation work and critical nucleus size increase, qualitatively in agreement with predictions. At large supersaturations, the impact of the flow on the nucleation kinetics is minimal, consistent with our previous study~\cite{richard2015role}. In appendix~\ref{ap:bdmd}, we compare Brownian dynamics simulations with our earlier work employing molecular dynamics and show that both dynamics can be mapped onto each other. It shows that the local dynamical rule has little effect on the nucleation kinetics (at least in the linear response regime). It is worth mentioning that the same observation has been made for the crystallization of soft spheres interacting via the Yukawa potential \cite{blaak2004crystal}. Continuing to increase the shear stress inside the liquid, we observe a deviation from the linear scaling.

\begin{figure}[t]
  \includegraphics[scale=1.0]{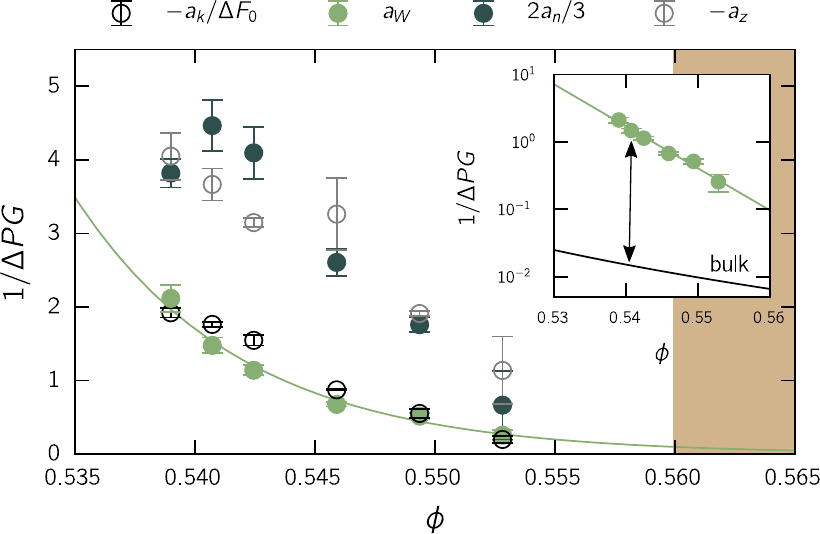}
  \caption{\textbf{Testing CNT predictions.} Relative test of response coefficients $\{-a_k/\Delta F_0,a_W,2a_n/3,-a_z\}$ as a function of the packing fraction $\phi$. The inset shows $a_W$ compared with the bulk prediction $a_W^\infty=1/(G_\infty\Delta P_\infty)$, $G_\infty$ and $\Delta P_\infty$ being the bulk solid shear modulus and the bulk pressure difference, respectively. The green line is an exponential decay.}
  \label{fig:cnttest}
\end{figure}

Linear regressions of the data in Fig.~\ref{fig:rate} gives us access to the response coefficients $\{a_k,a_W,a_n,a_z\}$ as a function of the packing fraction $\phi$. Again, we make explicit the difference between the $a_N$ and $a_Z$ derived from the thermodynamic modeling and the computed $a_n$ and $a_z$ based on the local bond order. In Fig.~\ref{fig:cnttest}, we plot $-a_k/\Delta F_0$, $a_W$, $2a_n/3$, and $-a_z$, which should collapse onto a single curve following the CNT prediction in Eq.~\eqref{eq:coef2}. We first observe that $-a_k/\Delta F_0$ does indeed fall onto $a_W$, indicating that the nucleation rate can be reasonably modeled by the increase of the nucleation work. Interestingly, we find $2a_n/3$ and $-a_z$ collapsing on top of each other, but with values about two times larger than $a_W$. A discrepancy between $a_W$ and $2a_n/3$ must stem from a non-linear relation between $N_c$ and $n_c$ since a simple rescaling would leave $a_n$ invariant. Such a non-linear relation can be attributed to the fact that we are considering critical droplets composed of hundred particles and less, which are thus mainly composed by particles located at the interface where the identification of ``solid-like'' particles is somewhat ambiguous.

Finally, we provide in the inset of Fig.~\ref{fig:cnttest} a comparison between $a_W$ and $a_W^\infty=1/(G_\infty\Delta P_\infty)$. The latter assumes that critical droplets have the same density $\rho_s^\infty$ than a bulk solid crystal at the same ambient liquid chemical potential $\mu_l$. We take the pressure difference $\Delta P_\infty$ from the bulk equations of state and $G_\infty$ from our parametrization of the shear modulus shown in Fig.~\ref{fig:visco}(b). We find that both $a_W$ and $a_W^\infty$ decrease exponentially as a function of $\phi$, which is consistent with the sharp increase of the shear modulus beyond the melting point [cf. Fig.~\ref{fig:visco}(b)]. The main observation, however, is that $a_W$ is about two orders of magnitude larger than the bulk prediction $a_W^\infty$. Although it is known that properties of small droplets deviate from bulk quantities, it is unlikely that such a gap can be explained by a hundred times smaller effective shear modulus. In the following, we will present a method to directly access the stress $\sig_s$ and strain $\gamma_s$ of critical droplets and demonstrate that the corrected response coefficient $a_W$ agrees with the CNT prediction.

\subsection{Seeding of droplets under shear}

\begin{figure}[b!]
  \includegraphics[scale=1.0]{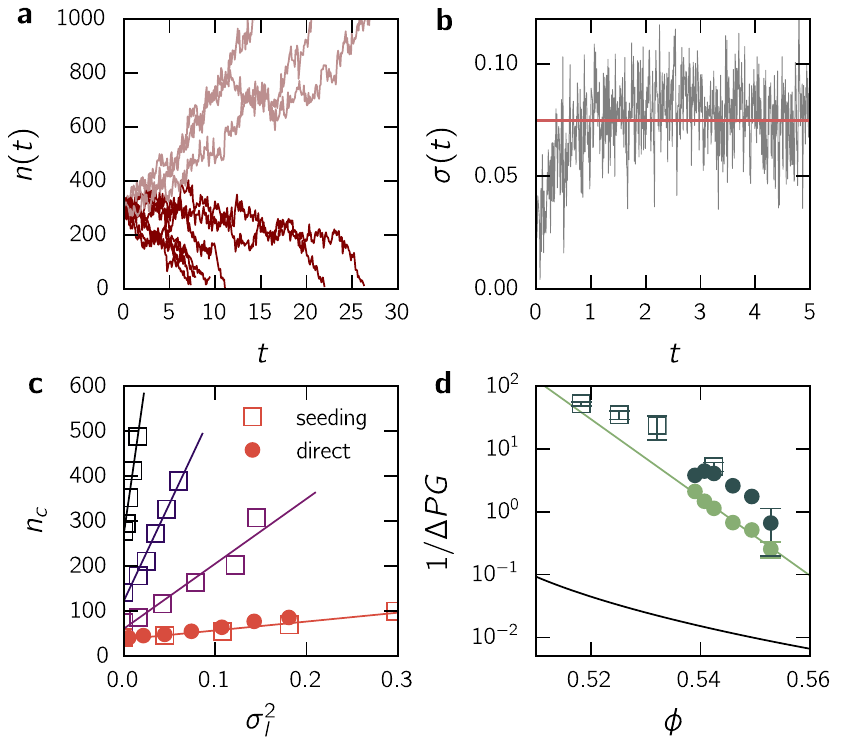}
  \caption{\textbf{Seeding of droplets under shear.} (a)~Evolution of the droplet size for 10 independent fleeting trajectories starting from a seed with $n\simeq 300$ for $\phi\simeq 0.518$ and $\dot\gamma\simeq 0.036$. (b)~Evolution of total shear stress in the system. The red lines corresponds to our parametrization $\sigma_l(\dot\gamma,\rho_l)=\eta(\dot\gamma,\rho_l)\dot\gamma$ [cf. Fig.~\ref{fig:visco}(a)]. (c)~Critical droplet sizes $n_c$ as a function of $\sigma_l^2$. Lines are linear regressions. (d)~Response coefficients $a_W$ and $2a_n/3$ as a function of the packing fraction $\phi$. Empty and filled symbols are from seeded and direct simulations, respectively.}
  \label{fig:seeding}
\end{figure}

To generate configurations of large critical droplets we now turn to a different type of simulations, namely the seeding of droplets~\cite{sanz2013homogeneous,espinosa2016seeding,richard2018crystallization2}. Starting configurations are prepared in the same way as described in our previous work~\cite{richard2018crystallization2}. To avoid finite size effects, we now study larger systems with $N=40,000$ particles. For each packing fraction, strain rate, and seed size, we generate $20$ fleeting trajectories that are terminated when either $n<n_a=10$ or $n>n_b=1000$ is fulfilled, see Fig.~\ref{fig:seeding}(a). We then compute the probability $P_B$ that a run crosses $n_b$ without coming back to $n_a$, which gives us an estimate for the critical nucleus for which $P_B\simeq1/2$ (cf. Eq.~\eqref{eq:pb}). As a consistency check, we show in Fig.~\ref{fig:seeding}(b) the evolution of the total shear stress in the system. We find $\sigma$ reaching a steady-state value for $t\simeq0.5-1\tau_B$, which is a negligible relaxation time compared with the typical fleeting time. Hence, our seeding preparation does not alter the extraction of $n_c$. In Fig.~\ref{fig:seeding}(c), we show our determination of the critical nucleus size $n_c$ as a function of the square of ambient shear stress $\sigma_l^2$. As found in Sec.~\ref{sec:coef}, we observe a linear behavior at small driving and we recover for the largest packing fraction ($\phi=0.542$) our previous results from the direct (unseeded) simulations. Linear regressions allow us to extract new estimates for $a_n$ until $\phi\simeq0.52$ and to compare them with $a_W$ and $a_W^\infty$. Clearly, no change of behavior can be seen although we are now probing critical droplets composed of several hundred particles.

\subsection{Extracting local shear stress and strain}

We are now proposing a methodology to access the solid density $\rho_s$, shear stress $\sigma_s$, and strain $\gamma_s$ of finite droplets, which ultimately will give us an estimate for the shear modulus $G$. To gather the necessary statistics, in this subsection we consider not only critical droplets but all droplets of a given size $n$.

\subsubsection{Local density and shear stress}

\begin{figure}[t!]
  \includegraphics[scale=1.0]{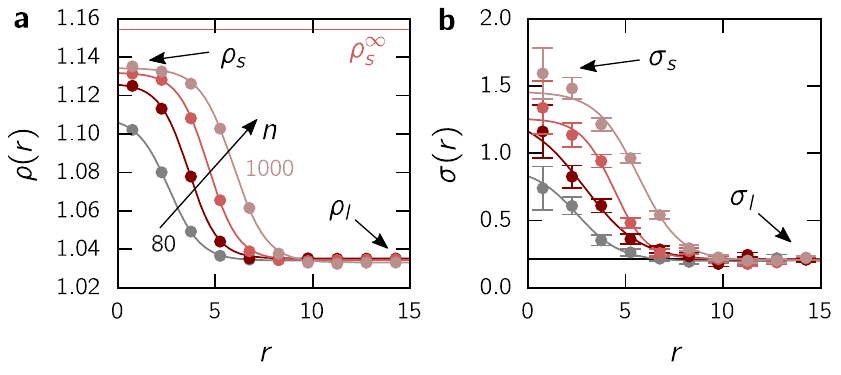}
  \caption{\textbf{Density and stress profiles.} (a)~Average radial density distribution $\rho(r)$ with respect to the center of mass of droplets for $\phi\simeq0.542$, $\dot\gamma\simeq0.036$ and nucleus sizes $n=80,250,500,$ and $1000$ from black to light orange. (b)~Average stress profile $\sigma(r)$. The horizontal dashed line indicates our parametrization of the liquid shear stress [cf. Fig.~\ref{fig:visco}(a)].}
  \label{fig:profiles}
\end{figure}

Using configurations generated via the seeding method, we are now able to extract density and stress profiles. To do so, we define for each particle $i$ its \emph{microscopic} local shear stress $\sigma_i$ by~\cite{egami1980structural}
\begin{equation}
  \hat\sigma_i=\frac{1}{2v_i}\sum_{j\ne i} \frac{x_{ij}y_{ij}}{r_{ij}} u'(r_{ij}),
\end{equation}
where $v_i$ is the Voronoi volume of the particle $i$. We can evaluate the average shear stress inside a given sub-volume $C$ via a weighted sum as
\begin{equation}
  \label{eq:sigsub}
  \sigma_\text{sub}=\frac{1}{\sum_{i\in C}v_i}\sum_{i\in C}v_i\hat\sigma_i,
\end{equation}
which we use to compute the radial stress profile $\sigma(r)$ with bin width $\Delta r=1.5d$ by measuring the particle positions with respect to the center of mass of the droplet. The radial density profile $\rho(r)$ follows simply from counting the number of particles in each bin. Profiles at a particular size $n$ are computed from a minimum of $400$ configurations and are taken with the criteria $n_i\in[n-5\%,n+5\%]$, $n_i$ being the size of the largest droplet in the system. In Fig.~\ref{fig:profiles}(a), we show such density profiles for $\phi\simeq0.542$ and nucleus sizes $n=80,250,500,$ and $1000$. Note that results for $n=80$ are extracted from configurations generated by direct simulations. As already discussed extensively in our previous work~\cite{richard2018crystallization2}, we observe a gradual increase of the density inside the solid droplet as its size grows. Profiles can be well modeled by the mean-field expression
\begin{equation}
  \rho(r)=\frac{\rho_l+\rho_s}{2}+\frac{\rho_l-\rho_s}{2}
  \tanh \left(\frac{r-R_0}{w}\right),
  \label{eq:rho}
\end{equation}
where $\rho_s=\rho(0)$ is the density at the center of the solid droplet, $\rho_l$ is the density of the surrounding liquid, $R_0$ is the radius at half maximum (different from $R_\ast$), and $w$ the interfacial width.

For $n=80$, we find a solid density $\rho_s\simeq1.11$ which would correspond (assuming bulk behavior) to a shear modulus $G\simeq66$. On the other hand, the bulk equation of state yields $\rho_s^\infty\simeq1.154$ and thus a significantly larger shear modulus $G_\infty\simeq 99$. This result marks our first finding to explain the discrepancy between $a_W$ and $a_W^\infty$. A smaller solid density results on one side in a smaller pressure difference between the two phases, thus decreasing the nucleation driving force $\Delta P$. On the other hand it results in a smaller shear modulus, increasing the elastic contribution to the nucleation work.

In Fig.~\ref{fig:profiles}(b), we show shear stress profiles from the same set of data for several droplet sizes $n$. We find that stress values at the center of droplets are about $3$ to $7$ times larger than the ambient shear stress $\sigma_l=\eta\dot\gamma$, to which all stress profiles decay. As observed in our density profiles, we find a gradual increase of the solid stress $\sigma_s=\sigma(0)$ as a function of the nucleus size $n$. This change is consistent with the increase of the density at the interface that locally increases the viscosity and thus the shear stress acting upon the nucleus. In fact, stress profiles can be well fitted by the same mean-field expression Eq.~\eqref{eq:rho} as for the density profiles. This result marks our second finding, namely that the elastic work to form a droplet is going to be much larger than the prediction made using $\sigma_l$ in Eq.~\eqref{eq:newwork} (already about $10$ times larger for $n=80$). It also explains why one cannot extract the response coefficient by plotting nucleation works against $\sigma_l^2$.

\subsubsection{Local shear strain}

\begin{figure}[t!]
  \includegraphics[scale=1.0]{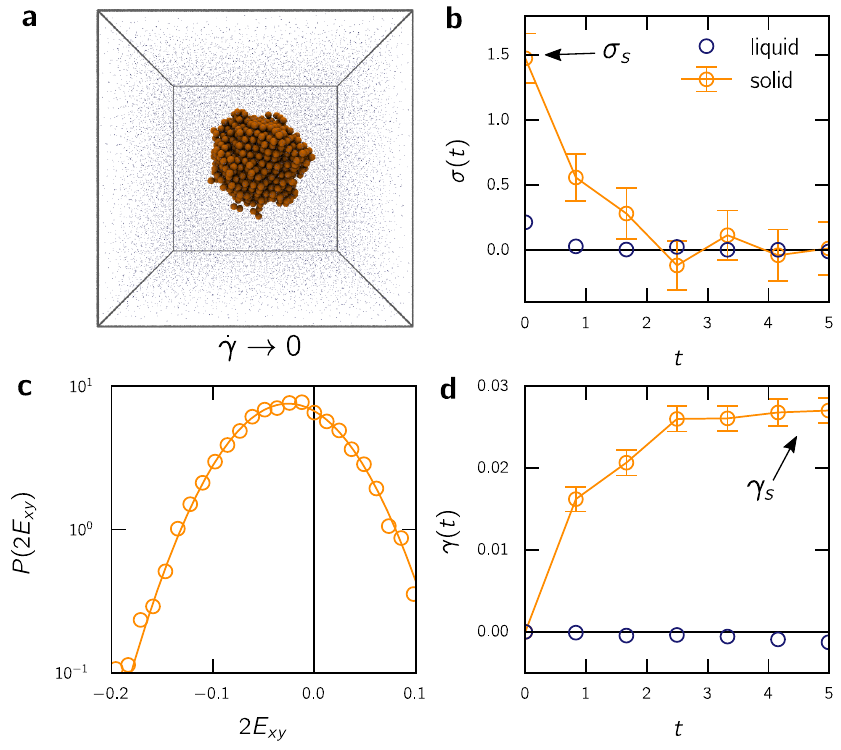}
  \caption{\textbf{Solid strain reconstruction.} (a)~Snapshot of a configuration with a strained droplet generated by the seeding method with $n=1000$ at $\phi\simeq 0.542$ and $\dot\gamma\simeq0.036$. (b)~Evolution of the average solid and liquid stress after switching off the shear flow. (c)~Probability distribution of two times the off-diagonal of the Green tensor $2E_{xy}$ for $t=5\tau_B$. The solid line is a normal distribution. (d)~Evolution of the average solid and liquid strain after switching off the imposed flow.}
  \label{fig:strain}
\end{figure}

Having evaluated the shear stress inside small droplets, we would like to characterize which elastic deformation ``solid-like'' particles have undergone. It is important to point out that there is not a unique and well-defined way to estimate local strains~\cite{falk1998dynamics,goldhirsch2002microscopic,graner2008discrete,tsamados2009local}. Here, we adopt a method developed in the context of the rheology of glasses~\cite{falk1998dynamics}, which is based on an adjustable strain tensor that fits best the actual particle displacements over some time interval. In our methodology, we harvest configurations with strained droplets, see snapshot in Fig.~\ref{fig:strain}(a), and switch off the imposed shear flow. Running short trajectories from these configurations, we can follow the elastic relaxation of the shear stress inside the solid phase and fit the local strain that particles undergo to release such a stress. In practice, we select a minimum of $400$ configurations with the same criteria as used for the density profiles. We run from them one trajectory of length $5\tau_B$. In Fig.~\ref{fig:strain}(b), we show for $n=1000$ the average solid stress taken for $r<2d$ and ambient liquid stress taken for $r>3R_0$. We find that the stress, both in the solid and liquid, relaxes fast with relaxation times of less than $3\tau_B$ and $1\tau_B$, respectively.

During each run, we evaluate the local displacement of a particle $i$ between a reference configuration at the starting time $t_0$ and a time $t$ through the deviatoric strain~\cite{falk1998dynamics} defined as 
\begin{equation}
  D^2_\text{min}(i,t_0\to t) = \sum_{j=1}^{n}[(\vec{r}_j(t)-\vec{r}_i(t))-\vec{D}\times(\vec{r}_j(t_0)-\vec{r}_i(t_0))]^2.
\end{equation}
Here, the sum runs over the $n$ closest neighbors of particle $i$, which are determined through a Voronoi tessellation at $t_0$. The deformation tensor $\vec{D}$ is determined by minimizing $D^2_\text{min}$. Because droplets can undergo body rotations, we evaluate the symmetric Green strain tensor $\vec{E} = \frac{1}{2}(\vec{D}^T\vec{D}-\vec{1})$. Finally, we can inspect the off-diagonal $E_{xy}$ to quantify shear deformations. In Fig.~\ref{fig:strain}(c), we plot the distribution $P(2E_{xy})$ at $t=5\tau_B$ for particles within a sphere of a radius $R=2d$ from the center of mass of the droplet. We observe a Gaussian distribution centered at negative values indicating that particles have undergone a reverse shear transformation with average strain $\gamma=-2\mean{E_{xy}}$. In Fig.~\ref{fig:strain}(d), we show the relaxation of the average strain in both the solid and liquid using the same criteria as for the stress shown in Fig.~\ref{fig:strain}(b). The strain in the solid phase follows the stress decay and reaches a plateau for times $t>3\tau_B$, yielding an estimate for the solid strain $\gamma_s$. As expected, the strain experienced by liquid particles is negligible.

\subsection{Shear modulus}

\begin{figure}[b!]
  \includegraphics[scale=1.0]{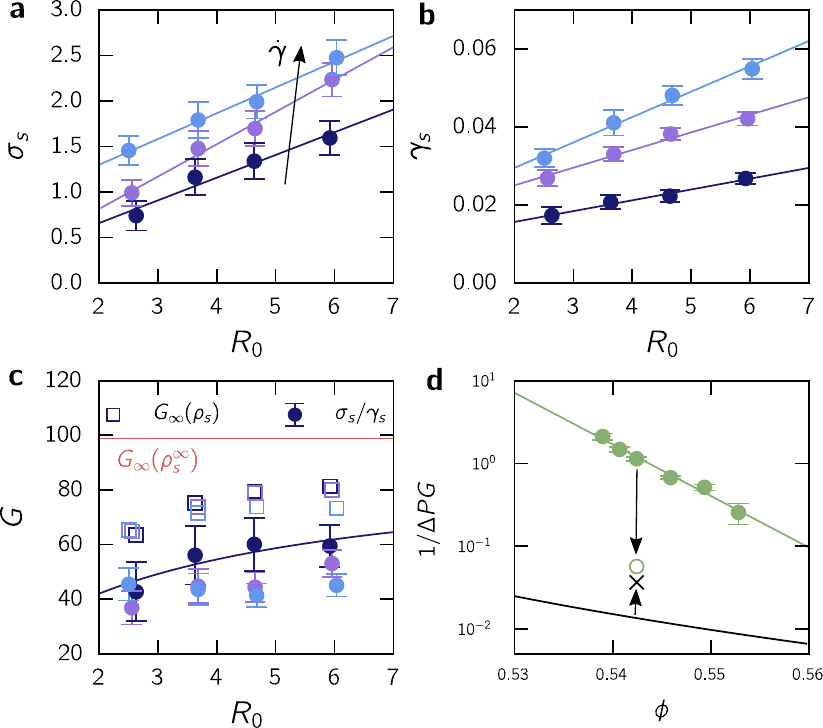}
  \caption{\textbf{Shear modulus.} (a)~Solid shear stress and (b)~strain as a function of the droplet radius $R_0$ for strain rates $\dot\gamma=0.036, 0.06$, and $0.084$. (c)~Shear modulus $G=\sigma_s/\gamma_s$ (filled symbols) of small droplets compared with $G_\infty(\rho_s)$ (empty symbols) employing the actual droplet density $\rho_s$. The red horizontal line indicates the bulk estimate $G_\infty(\rho_s^\infty)\simeq 99$. All simulations in this figure are performed at $\phi\simeq0.542$. (d)~Corrected response coefficient $a_W$ (empty disk, obtained in analogy with Fig.~\ref{fig:rate} but plotting \emph{vs.} $\sig_s^2$) and $a_W^\text{CNT}$ (cross, employing $G\simeq65$ and $\Delta P\simeq0.60$)}
  \label{fig:modulus}
\end{figure}

We collect the shear stress $\sigma_s$ [Fig.~\ref{fig:modulus}(a)] and strain $\gamma_s$ [Fig.~\ref{fig:modulus}(b)] at packing fraction $\phi\simeq0.542$ and three values $\dot\gamma\simeq0.036,0.06,0.084$ of the strain rate. The radius ranges from $2.5$ to $6$ particle diameters. As expected for a fixed droplet size, increasing $\dot\gamma$ results in an increase of both $\sigma_s$ and $\gamma_s$. Moreover, we find that both quantities scale approximately linearly with the droplet radius, which is a direct consequence of the increasing droplet density.

We can now extract estimates for the shear modulus of small droplets via $G=\sigma_s/\gamma_s$. In Fig.~\ref{fig:modulus}(c), we compare $G$ with both the bulk prediction $G_\infty(\rho_s^\infty)$ and the bulk prediction at the actual droplet density, $G_\infty(\rho_s)$. We find that both values increase slightly with $R_0$ and follow closely the trend of $G_\infty(\rho_s)$ although being $30-40\%$ smaller. For small droplets, shear moduli are found to be close to $G\simeq40$ and thus significantly smaller than the bulk value $G_\infty(\rho_s^\infty)\simeq100$.

Having extracted the stress inside droplets, for one density (using droplets composed of $80$ particles at $\dot\gamma\simeq0.084$) we invert $W(\sigma_s)=\Delta F_0(1+a_W\sigma_s^2)$ to get a new estimate for $a_W$. In addition, we also correct our previous estimate of $a_W^\text{CNT}$ using $G$ and $\Delta P$ instead of $G_\infty$ and $\Delta P_\infty$, respectively. Here, the corrected pressure difference $\Delta P\simeq0.8\Delta P_\infty$ is taken from Ref.~\citenum{richard2018crystallization2}. In Fig.~\ref{fig:modulus}(d), we compare our previous estimates with the corrected response coefficients. We now find a much better agreement between the extracted $a_W$ and the CNT prediction $a_W^\text{CNT}$.

\subsection{Elastic work}

\begin{figure}[b!]
  \includegraphics[scale=1.0]{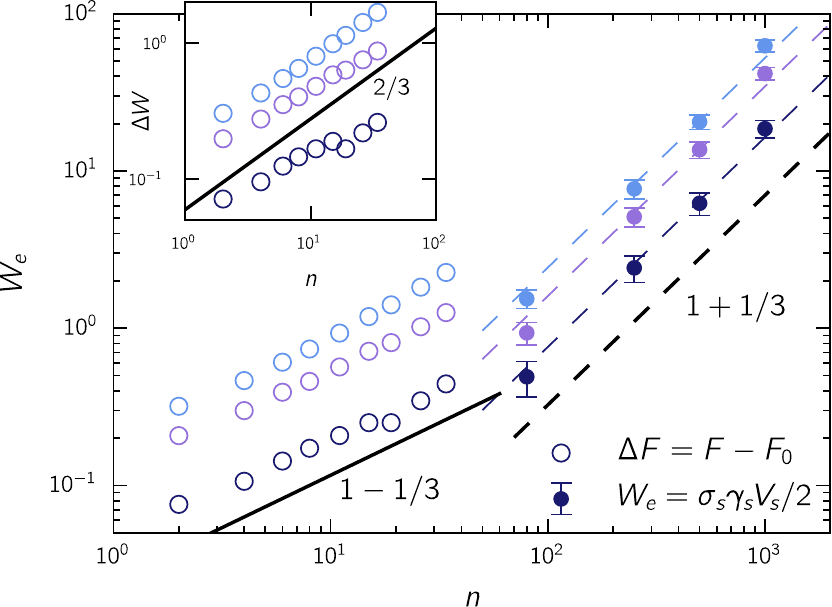}
  \caption{\textbf{Elastic work.} Empty symbols are the excess ``free energy'' $\Delta F$ [Eq.~\eqref{eq:F:exc}] and filled symbols corresponds to the direct evaluation [Eq.~\eqref{eq:elwork:2}] of the elastic work $W_e$ as a function of nucleus size $n$ at $\phi\simeq0.542$ for strain rates $\dot\gamma\simeq0.036,0.06,0.084$ (bottom to top). Dashed lines model the elastic work as $W_e(n)\sim n^{1+1/3.}$. The inset shows the difference $\Delta W(n;\gdot)=\Delta F(n;\gdot)-W_e(n;\gdot)$, where $W_e(n;\gdot)$ is taken from an extrapolation of $W_e$ to smaller sizes $n$.}
  \label{fig:elwork}
\end{figure}

Since for small droplets we have access to the full effective free energy $F(n;\gdot)$, we can extract the excess
\begin{equation}
  \label{eq:F:exc}
  \Delta F(n;\gdot) = F(n;\gdot) - F_0(n) = W_e(n;\gdot) + \Delta W
\end{equation}
due to the shear flow for all droplet sizes (not only critical droplets). We split this excess into the elastic work
\begin{equation}
  \label{eq:elwork:2}
  W_e(n;\gdot) = \frac{1}{2}\sig_s(n)\gamma_s(n)V_s(n)
\end{equation}
eliminating $G$ from Eq.~\eqref{eq:elwork} and further contributions $\Delta W$ that are not captured by the theory presented in Sec.~\ref{sec:theory}. In Fig.~\ref{fig:elwork}, we plot $\Delta F(n)$ together with the direct evaluation of $W_e$ for large droplets obtained from the seeded simulations, where we employ $V_s=\frac{4\pi}{3}R_0^3$ approximating the radius of tension $R_s$ by $R_0$. While covering different sizes, we find that both estimates follow the same trend although $\Delta F$ clearly overestimates the true elastic work $W_e$. This strongly indicates that the elastic contribution does not capture the entire change in the nucleation work for sheared melts and a positive term $\Delta W>0$ is still missing.

To gain further insight, we consider the scaling of $W_e$ with the number $n$ of solid particles. Employing bulk quantities, the stress $\sigma_s$, strain $\gamma_s$, and the density $\rho_s$ would be independent of $n$, which would result in a scaling $W_e\sim n$ since $V_s\sim n/\rho_s$. In contrast, we observe $W_e\sim n^{1+1/3}$, which agrees with the linear increase of $\sigma_s$, $\gamma_s$, and $\rho_s$ with respect to $R_0$ as seen in Fig.~\ref{fig:modulus} and Fig.~\ref{fig:profiles}(a). Interestingly, we observe a different scaling for $\Delta F\sim n^{2/3}$, suggesting a work that is dominated by the droplet area. We confirm this result by extrapolating $W_e$ to small sizes and plotting the difference $\Delta W=\Delta F-W_e\sim n^{2/3}$ [inset of Fig.~\ref{fig:elwork}], which indeed behaves as a surface term. Moreover, we find $\Delta W\simeq k_BT$ for $n\sim40$, which can already result in a significant change in the nucleation kinetics. For large droplets, we expect a crossover to a regime in which the elastic volume term $W_e$ will dominate the excess $\Delta F$.


\section{Discussions}

Performing direct and seeded simulations of sheared hard spheres, we have demonstrated that an elastic contribution plays an important role in the change of the nucleation work for sheared liquids. Including this contribution, CNT makes near-quantitative predictions as long as one takes into account that (i)~the density inside droplets does not reach its bulk value, $\rho_s<\rho_s^\infty$, and (ii)~the solid shear stress is significantly larger than the ambient liquid shear stress.

Extracting independently the work difference $\Delta F=F-F_0$ from free energy calculations and the elastic work $W_e$ from computing the local stress and strain of small droplets, we find the two estimates to be consistent with each other, although they do not agree perfectly. We trace back this discrepancy to an additional work $\Delta W$ that the system has to spend in order to form a droplet. This suggests that the overall nucleation work to form a solid nucleus of size $n$ takes the form
\begin{equation}
  \Delta F(n;\gdot) = F_0(n) + W_e(n;\gdot) + \Delta W(n;\gdot).
\end{equation}
As discussed in Ref.~\citenum{mura2016effects}, one expects that shear induces a global deformation of the nucleus into a more ellipsoidal shape, which we have neglected here. To lowest order, one expects an increase $\delta A$ of the droplet surface that scales as $\delta A\sim\dot\gamma^2$. Such an increase could explained why $\Delta W\sim n^{2/3}>0$, although one should bear in mind that it could also stem from an increase of the interfacial tension $\Gamma$. In fact, experimentally it has been observed for a colloidal gas-liquid interface that shear suppresses capillary waves and tends to increase $\Gamma$~\cite{derks2006suppression}. Here, the microscopic picture is a local erosion at the interface, where particles rattle due to the stress and eventually dissolve into the liquid.

Here we have focussed on the pressure difference $\Delta P$ as the natural driving force of crystallization. For an incompressible solid, one can rewrite the driving force as the (absolute) difference of chemical potential $|\Delta \mu|=\mu_l-\mu_s$ between the solid and the liquid phase at the same ambient pressure $P_l$. For a quiescent liquid, increasing $P_l$ beyond the coexistence pressure $P_\text{coex}$, the two branches $\mu_l(P)$ and $\mu_s(P)$ move apart from each other increasing $|\Delta\mu|$ and thus leading to a faster nucleation process. Hence, to explain the suppression of nucleation under shear using the same framework one would require a non-equilibrium liquid chemical potential that becomes smaller. As discussed by Butler and Harrowell~\cite{butler2002factors,butler2003simulation}, there is no consistent definition of a non-equilibrium chemical potential that would predict such a reduction.

Indeed, we provide new numerical evidence in Appendix~\ref{ap:chem} that confirms the opposite scenario with an increase of $\mu_l$ as a function of $\dot\gamma$. To do so, we have employed the fast growth method~\cite{hendrix2001fast} to extract the insertion work $w_{\text{ex}}$ needed to place a particle in a (dense) melt. For a quiescent liquid this work reduces to the excess chemical potential $\mu_l^{ex}=\mu_l-\mu_l^{id}$, with $\mu_l^{id}$ the chemical potential of an ideal gas. Consistent with the flow symmetry, we find in the linear regime that $w_{\text{ex}}$ increases with $\dot\gamma^2$. Identifying this work with a non-equilibrium liquid chemical potential, it moves away from the solid branch and increases the driving force $|\Delta \mu|$, which is inconsistent with the observed suppression of nucleation.


\section{Conclusions}

In this paper, we have tested a possible extension of classical nucleation theory to model the crystallization of sheared liquids. We have demonstrated that one of the key ingredient to model the change of nucleation work as a function of the imposed strain rate $\dot\gamma$ is to take into account the elastic deformation of the newborn nucleus. Such an extension can already predict, on a qualitative level, many observations made in previous simulations, namely the drop of the nucleation rate and the increase of both the nucleation work and the critical nucleus size as a function of $\dot\gamma$.

To obtain quantitative predictions, one needs to evaluate the elastic work $W_e$, which contains the shear stress $\sigma_s$ inside the droplet and its shear modulus $G$. These two quantities are unknown, hence one has to employ approximations. As a first step, one might try to: (i)~equating the solid shear stress with the ambient liquid stress $\sigma_l=\eta\dot\gamma$ and (ii)~to treat small droplets as bulk phases and employ the bulk shear modulus $G_\infty$ of a macroscopic solid. In Sec.~\ref{sec:coef}, we have demonstrated that these approximations do not lead to satisfactory results.

In the second part of this paper, we have employed a seeding method that generates configurations with larger droplets under shear, from which we have extracted the local solid density $\rho_s$, the shear stress $\sigma_s$, and the strain $\gamma_s$. Our first finding is a direct consequence of what is seen in quiescent crystallization, namely that the density at the center of droplets does not reach its bulk value, leading to an effective smaller shear modulus compared with the bulk value. Our second observation is the sharp increase of shear stress inside the nucleus compared with the ambient hydrodynamic stress $\sigma_l$. Employing the actual values $\sigma_s$ and $G$, we indeed find a much better agreement between numerics and CNT.

Finally, we have highlighted the presence of another term entering the nucleation work, $\Delta W$, and have argued that it arises from a positive contribution to the interfacial work. Due to scaling reasons, this term dominates the work for small nuclei but should become negligible compared to the bulk elastic work for large droplets. How the interfacial tension $\Gamma$ changes when the liquid layer at contact with the droplet is sheared remains to be investigated.

In the present manuscript, we have not addressed the regime of deeply supercooled liquids, in which the change of the nucleation work competes with the enhanced dynamics that increases the kinetic prefactor present in the nucleation rate. A natural extension of this work includes quantifying how the diffusion in droplet size space varies as a function of the liquid density and strain rates.


\begin{acknowledgments}
  We gratefully acknowledge ZDV Mainz for computing time on the MOGON supercomputer. We acknowledge financial support by the DFG through the collaborative research center TRR 146.
\end{acknowledgments}

\appendix

\section{Brownian dynamics vs. molecular dynamics}
\label{ap:bdmd}

\begin{figure}[t]
  \includegraphics[scale=1.0]{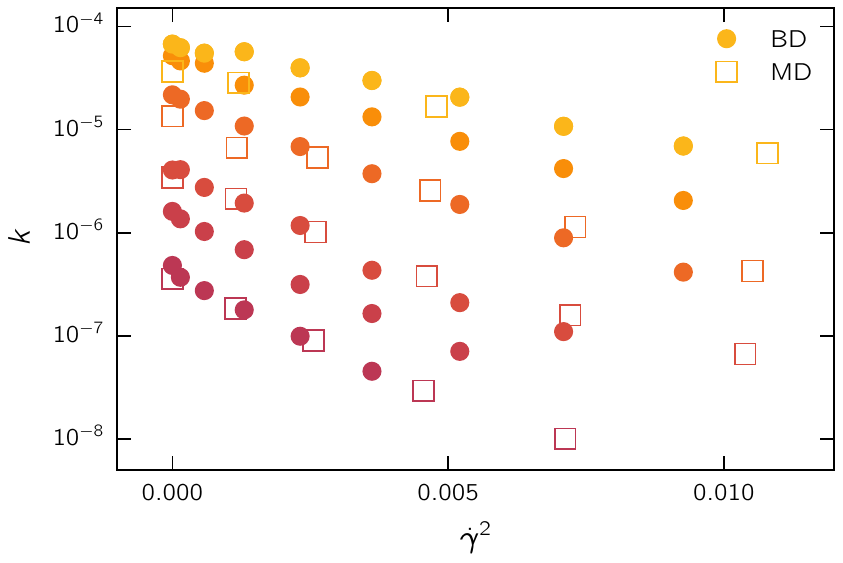}
  \caption{\textbf{Brownian dynamics vs. molecular dynamics.} Comparison between the nucleation rates $k$ extracted from BD and MD simulations as a function of the square of the strain rate $\dot\gamma^2$.}
  \label{fig:bdmd}
\end{figure}

We perform molecular dynamics (MD) simulations in the NVT ensemble using the Lowe-Anderson thermostat~\cite{Lowe06} has in our previous work, Ref.~\citenum{richard2015role}. In these simulations, time is measured in units of $\sqrt{m\alpha^2/k_BT}$, with $m$ being the mass of particles. A velocity Verlet integrator is used with a time step $\Delta t=0.004$. The coupling between the system and the thermostat is regulated by the bath collision frequency $\Gamma=10$. Rates are again extracted via the plateau of the mean first passage time. We then rescale the MD nucleation time in units of the Brownian time $\tau_B^{MD}=d^2/D_0$, where
\begin{equation}
  D_0 = \frac{3}{8\sqrt{\pi}}\left(\frac{k_BT}{m}\right)^{1/2}\left(\frac{1}{\rho d^2}\right)
\end{equation}
is the bare self-diffusion coefficient taken from the Chapman-Enskog kinetic theory of gases~\cite{chapman1990mathematical}. It allows a rescaling of the nucleation rate $k$ as well as the strain rate $\dot\gamma$. In Fig.~\ref{fig:bdmd}, we compare nucleation rates extracted from BD and MD at various packing fractions. We find a good agreement between the two dynamics indicating the consistency between this work and our previous study in Ref.~\citenum{richard2015role}.

\section{Non-equilibrium chemical potential}
\label{ap:chem}

\begin{figure}[t]
  \includegraphics[scale=1.0]{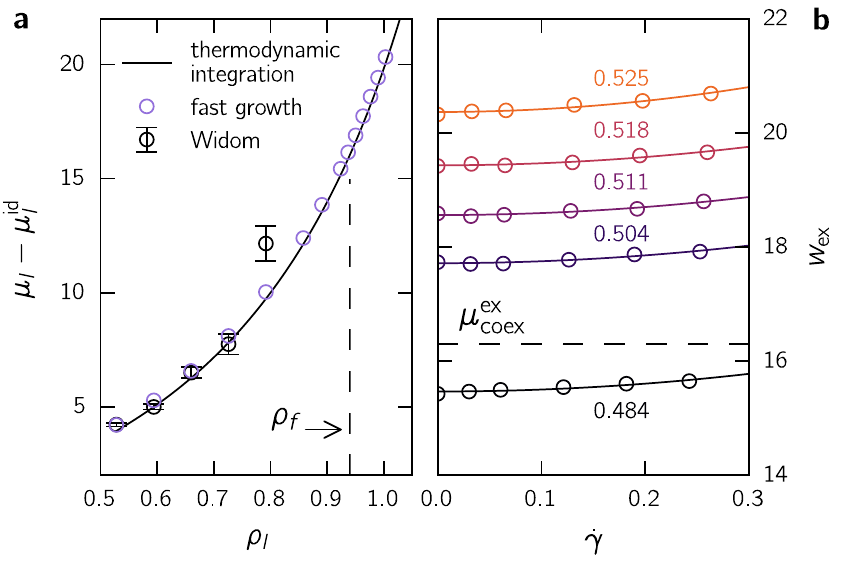}
  \caption{\textbf{The fast growth method.} (a)~Excess chemical potential $\mu_l-\mu_l^{\text{id}}$ as a function of the liquid density $\rho_l$. Shown are results from the fast growth method (purple symbols), Widom insertion method (black symbols), and thermodynamic integration (black solid line). (b)~Insertion work $w_{\text{ex}}$ as a function of the strain rate $\dot\gamma$ for various packing fractions $\phi$. Lines are quadratic fits.}
  \label{fig:chem}
\end{figure}

The fast growth method~\cite{hendrix2001fast} is based on the Jarzynski equality
\begin{equation}
  \exp(-\Delta\mathcal F) = \langle\exp(-\beta w_\tau)\rangle,
\end{equation}
relating the free energy difference $\Delta\mathcal F$ between two states $A$ and $B$ to the distribution of work $w_\tau$ performed on the system to move it from $A$ to $B$ within time $\tau$. Here, we evaluate the free energy difference between a liquid composed of $N$ and $N+1$ particles at a fixed volume $V$, which is nothing than the microscopic definition of the chemical potential $\mu_l$. The work $w_\tau$ is computed from a discrete protocol, where we progressively switch on the interaction between a tagged particle and the surrounding fluid through changing the parameter $\lambda$ from $0$ to $1$. The work reads
\begin{equation}
  w_\tau = \sum_{t=0}^{\tau-1}[H_{\lambda_{t+1}}(\omega_t)-H_{\lambda_{t}}(\omega_t)],
\end{equation}
where $H_{\lambda_{t}}(\omega_t)$ is the Hamiltonian at time $t$ with microstate $\omega_t$ (position and velocities of all particles). In practice, we use the protocol $\lambda(t)=(t/\tau)^6$. In the limit of instantaneous switching, $\tau\to0$, one recovers the Widom insertion method~\cite{allen2017computer}. In Fig.~\ref{fig:chem}(a), we show the (excess) liquid chemical potential $\mu_l-\mu_l^{\text{id}}$ as a function of the density $\rho_l$ for $\tau=10^5\Delta t$ and we compare these new results with the thermodynamic integration used in Ref.~\citenum{richard2018crystallization}. We find a perfect agreement with our previous parametrization even beyond the freezing density $\rho_f$. In contrast, the Widom insertion method ($\tau\to0$) is only able to compute chemical potential differences that are below $\sim10k_bT$. Applying the fast growth method in a sheared liquid, we find a continuous increase of the insertion work $w_{\text{ex}}=-\ln(\langle\exp(-\beta w_\tau)\rangle)$ as a function of the strain rate $\dot\gamma$, cf. Fig.~\ref{fig:chem}(b). 


%

\end{document}